\documentclass[journal,twoside,web]{ieeecolor}
\usepackage{generic}
\usepackage{cite}
\usepackage{cuted}
\usepackage{amsmath,amssymb,amsfonts}
\usepackage{algorithm}
\usepackage{algpseudocode}
\usepackage{graphicx}
\usepackage{textcomp}
\usepackage{subfigure}
\usepackage{subcaption}
\usepackage[justification=centering]{caption}
\def\BibTeX{{\rm B\kern-.05em{\sc i\kern-.025em b}\kern-.08em
    T\kern-.1667em\lower.7ex\hbox{E}\kern-.125emX}}
\begin{document}
\title{A Group Consensus-Driven Auction Algorithm for Cooperative Task Allocation Among Heterogeneous Multi-Agents}
% \author{First A. Author, \IEEEmembership{Fellow, IEEE}, Second B. Author, and Third C. Author, Jr., \IEEEmembership{Member, IEEE}
% \thanks{This paragraph of the first footnote will contain the date on 
% which you submitted your paper for review. It will also contain support 
% information, including sponsor and financial support acknowledgment. For 
% example, ``This work was supported in part by the U.S. Department of 
% Commerce under Grant BS123456.'' }
% \thanks{The next few paragraphs should contain 
% the authors' current affiliations, including current address and e-mail. For 
% example, F. A. Author is with the National Institute of Standards and 
% Technology, Boulder, CO 80305 USA (e-mail: author@boulder.nist.gov). }
% \thanks{S. B. Author, Jr., was with Rice University, Houston, TX 77005 USA. He is 
% now with the Department of Physics, Colorado State University, Fort Collins, 
% CO 80523 USA (e-mail: author@lamar.colostate.edu).}
% \thanks{T. C. Author is with 
% the Electrical Engineering Department, University of Colorado, Boulder, CO 
% 80309 USA, on leave from the National Research Institute for Metals, 
% Tsukuba, Japan (e-mail: author@nrim.go.jp).}}

\author{Gang Wang, Hongfang Han, Xiaowei Liu, Hanfeng Jiang, Ming Zhang \thanks{Gang Wang is with Information School, Xi'an University of Finance and Economics, Xi'an, 710100 China (e-mail: gangw@xaufe.edu.cn).} \thanks{Hongfang Han is with Faculty of Intelligence Technology, Shanghai Institute of Technology, Shanghai, 201418 China. She is the corresponding author (e-mail: hanhf@sit.edu.cn).} \thanks{Xiaowei Liu and Hanfeng Jiang are with College of Intelligence and Computing, Tianjin University, Tianjin, 300350 China (e-mail: xiaoweiliu@tju.edu.cn, tjujhf@tju.edu.cn).} \thanks{Ming Zhang is with Faculty of Environment, Science and Economy, University of Exeter, Exeter, EX4 4QJ UK (e-mail: mz427@exeter.ac.uk).}}

\maketitle

\begin{abstract}
In scenarios like automated warehouses, assigning tasks to robots presents a heterogeneous multi-task and multi-agent task allocation problem. However, existing task allocation study ignores the integration of multi-task and multi-attribute agent task allocation with heterogeneous task allocation. In addition, current algorithms are limited by scenario constraints and can incur significant errors in specific contexts. Therefore, this study proposes a distributed heterogeneous multi-task and multi-agent task allocation algorithm with a time window, called group consensus-based heterogeneous auction (GCBHA). Firstly, this method decomposes tasks that exceed the capability of a single Agent into subtasks that can be completed by multiple independent agents. And then groups similar or adjacent tasks through a heuristic clustering method to reduce the time required to reach a consensus. Subsequently, the task groups are allocated to agents that meet the conditions through an auction process. Furthermore, the method evaluates the task path cost distance based on the scenario, which can calculate the task cost more accurately. The experimental results demonstrate that GCBHA performs well in terms of task allocation time and solution quality, with a significant reduction in the error rate between predicted task costs and actual costs.
\end{abstract}


\section{Introduction}
\label{sec:introduction}
\IEEEPARstart{I}{n} scenarios such as automated warehouses, multi-robots work together to fulfill a set of cargo orders, where the robots need to move to specified locations to pick up and then deliver the goods to designated positions. The system needs to assign tasks to the robots and plan their paths. This process is called multi-agent pickup and delivery (MAPD)\cite{ma2017lifelong}, as seen in Fig.1. MAPD encompasses two problems: multi-agent path finding (MAPF) and task allocation (TA). 1) The non-conflicting movement of agents to specified locations for picking up and delivering goods can be regarded as a MAPF problem; 2) Assigning tasks to be executed by all agents is a TA problem. Among them, TA is the process of optimally assigning tasks to agents based on their current states, aiming to minimize overall completion cost or time.

    \begin{figure*}[htbp]
    \centering
 \centerline{\includegraphics[width=0.73\textwidth]{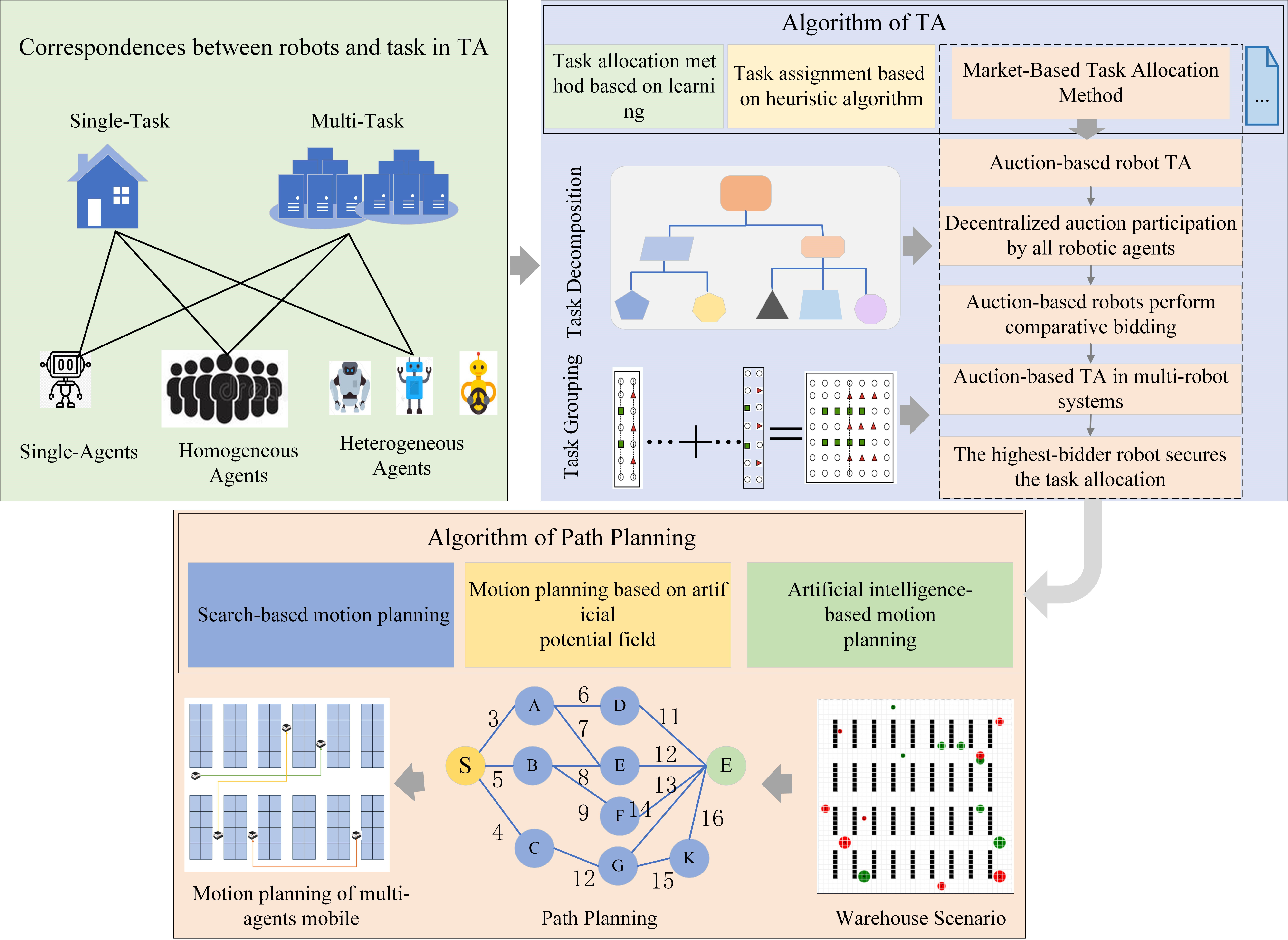}}
    \caption{Processing of Multi-Agent Pickup and Delivery}
    \label{fig1}
    \end{figure*}
    
In multi-robot task allocation, there are four types of correspondences between robots and tasks: single task-single robot, single task-multiple robots, multiple tasks-single robot, and multiple tasks-multiple robots\cite{gerkey2004formal}. Currently, most researchers have focused on the problem of multi-task single-robot task allocation. For example, Amanda et al. \cite{whitbrook2019addressing} proposed a robustness module to improve existing task allocation algorithms, enhancing the algorithm’s performance in uncertain environments.A few researchers have also explored the problem of multi-task multi-robot task allocation. Zhao et al.\cite{zhao2015heuristic} proposed a new heuristic distributed task allocation method, which allocates tasks by defining the significance and contribution of each task. Ayan et al. \cite{dutta2019correlation}employed a linear programming-based graph partitioning method and a region growth strategy for robot grouping and task allocation. Lu et al.\cite{lu2018multiple}proposed a multi-site swarm foraging algorithm, where robots move based on the perceived average task target location and can transport multiple targets back. 
While prior work has examined both multi-task multi-robot allocation or heterogeneous task allocation, studies investigating their integration known as heterogeneous multi-task multi-robot task allocation remain relatively limited. Heterogeneous task allocation is divided into two categories: 1) agents sharing the same type but with varying parameters, and 2) fundamentally different agent types. The first category primarily addresses the allocation of identical tasks among similar agents. The second category focuses on solving the problem of different types of agents performing different types of tasks, which requires the proper assignment of tasks to agents. Notably, some researchers have also considered the combination of these two types of heterogeneous task allocation problems, where different types of agents may have some functions in common, although the parameters of these functions may differ. In this context, Chen et al.\cite{chen2022efficient} proposed a new multi-objective ant colony optimization algorithm specifically for heterogeneous unmanned aerial vehicle task allocation.

Moreover, in scenarios such as warehouse systems, task allocation poses significant challenges. Current dynamic task allocation classified into four categories: market-based, optimization-based, behavior-based, and task-cluster-based methods. Among these, market-based task allocation has emerged as a particularly effective strategy \cite{schneider2017mechanism}. This strategy mimics the form of market transactions and can be implemented in both distributed and centralized structures. For example, Choi et al. \cite{choi2009consensus} proposed the consensus-based auction algorithm and the consensus-based bundle algorithm (CBBA), which integrate consensus-based algorithms with auction mechanisms, enabling operation in weak communication network conditions. Based on CBBA, Hunt et al. \cite{hunt2014consensus} subsequently proposed the consensus-based grouping algorithm (CBGA) to address multi-task coordination challenges.  As an advanced task allocation framework, CBGA demonstrates superior performance in comparative studies: it outperforms other market-based approaches in terms of both solution quality and system scalability. However, its consensus formation speed remains limited by both the agent population size and task complexity.

Additionally, unlike general task allocation that considers only a single target location, automated warehouse scenarios typically involve tasks with both pickup and delivery locations. This fundamental difference renders conventional task allocation algorithms inapplicable. While Euclidean distance is commonly employed as a path cost estimator to maintain algorithm generality, this approach can yield substantial estimation errors in structured environments with known layouts.
To address the limitations in the aforementioned research, this paper proposes an enhanced CBGA-based algorithm for solving heterogeneous multi-robot multi-task allocation problems in generalized MAPD. The proposed methodology integrates market-based and task-clustering approaches through three key innovations: First, we implement a task decomposition mechanism that breaks down large-scale tasks into executable sub-tasks for individual agents. A heuristic clustering method then groups these sub-tasks, which serves dual purposes: 1) Reducing the bidding workload for agents, thereby accelerating consensus formation. 2) Maintaining solution quality through optimized task grouping. Second, recognizing the regular shelf arrangements characteristic of MAPD environments, we develop a scenario-specific path cost estimation method. This approach leverages environmental structure to generate more accurate path cost predictions, significantly improving actual path cost reduction compared to conventional estimation techniques. Third, a computational experiment model is constructed to verify the performance of the task allocation algorithm proposed in this paper.
The proposed solution not only provides theoretical advancements in heterogeneous task allocation but also offers a practical methodology for addressing  generalized multi-agent pickup and delivery (GMAPD) challenges, thereby facilitating real-world MAPD implementations.

\section{BACKGROUND AND MOTIVATION}
 In this section, we focus on three key aspects: multi-agent systems (MAS), MAPD and computational experiment, where we systematically analyze the limitations of existing approaches and propose corresponding enhancements.

\subsection{Multi-Agent Systems}

An agent is defined as a computer system situated within an environment and capable of autonomously executing actions to achieve designated objectives \cite{wooldridge1995intelligent}. MAS has evolved in response to the growing demand for interconnected distributed systems, leading to landmark applications such as autonomous space probes \cite{muscettola1998remote} and intelligent air traffic control \cite{10.5555/60204.60208}. While early MAS were based on homogeneity assumptions implying consistent agent states and capabilities the increasing prevalence of heterogeneous collaboration scenarios has shifted research focus toward two categories of heterogeneous MAS: 1) Isomorphic dynamic nonlinear systems, in which agents are structurally identical but exhibit parametric nonlinear variations \cite{app11104345}. 2) Cross-modal heterogeneous systems, characterized by mismatched state-space or kinematic models \cite{sun2021mean}. Current research prioritizes coordination strategies for heterogeneous MAS—favoring output consensus over state consistency and explores task allocation approaches ranging from independent task cost optimization to collaborative task decomposition \cite{xue2016computational}. Nevertheless, a significant gap remains between theory and application, particularly in achieving unified multi-dimensional heterogeneity modeling and developing dynamic real-time coordination mechanisms for industrial deployment \cite{zhou2024hierarchical, yu2025unlocking}.

\subsection{Multi-Agent Pickup and Delivery}

Multi-agent pickup and pelivery (MAPD) consists of task allocation and MAPF. The MAPF problem, which has been proven to be NP-hard\cite{sharon2015conflict}, requires multiple agents to compute collision-free paths from their initial positions to designated target locations within a known environment. Various MAPF algorithms have been proposed in existing research, including search-based , prioritized planning, rule-based, and learning-based methods\cite{sharon2015conflict} \cite{li2022multi}\cite{sartoretti2019primal}. These approaches exhibit distinct advantages and limitations depending on application contexts. For example, search-based methods guarantee solution optimality \cite{wagner2011m} at the expense of substantial computational overhead, whereas learning-based methods demonstrate superior scalability for large-scale multi-agent systems\cite{sartoretti2019primal} but typically sacrifice solution optimality guarantees. The task allocation problem focuses on how to complete all tasks at the lowest cost or in the shortest time. In MAPD contexts, the inherent pickup-delivery duality of tasks necessitates specialized algorithmic adaptations\cite{xue2022computational}. Existing research has developed diverse allocation methodologies, notably including: market-based approaches and optimization-based techniques. However, these methods usually rely on Euclidean distance to estimate path costs, leading to substantial discrepancies between estimated and actual path costs\cite{wang2012multi}. Consequently, investigating generalized MAPD problems incorporating both heterogeneous agents and tasks holds considerable theoretical and practical significance.

\subsection{Computational Experiment}
In 2004, the term "computational experiment" was proposed and a methodological system of "artificial systems + computational experiments + parallel execution" was established. Modern computational experimentation has reached methodological maturity, covering: artificial society modeling, experimental system construction, experimental design, analysis, and validation \cite{xue2024computational1, xue2024computational2}. However, key challenges remain in model validation, experimental design, and real-world system mapping, necessitating further research.
In MAPD research, the computational experiment method provides a powerful tool for simulating and analyzing complex systems \cite{xue2018social, zhou2022sle2}. Current MAPF algorithms often assume homogeneous agents, ignoring their physical attributes and capabilities. This oversimplification results in suboptimal path planning performance in practical applications featuring heterogeneous robotic teams. Furthermore, traditional task allocation algorithms struggle to efficiently assign tasks to agents with varying capacities and requirements, especially in scenarios where tasks have multiple locations and complex constraints. These limitations highlight the need for improved algorithms that can effectively address the heterogeneity of agents and tasks, thereby enhancing the overall efficiency and applicability of MAPD solutions in complex environments.

\section{METHOD}

In this Section, we present a novel heterogeneous auction framework based on group consensus principles. As illustrated in Fig.1, the proposed methodology comprises four steps, which is represented below in detail.

\subsection{Step 1: The definition of heterogeneous multi-task multi-robot task allocation}
The input of the heterogeneous multi-task multi-robot task allocation problem in generalized MAPD is a triplet $\{G(V,E),Agents,Tasks\}$, where $G(V,E)$ is the map where the agent and tasks are located, V represents the set of positions,E indicates the set of paths traversed from one position to another. The purpose of the graph $G(V,E)$ is to estimate the task path costs more accurately. Agents are composed of $n$ Agent $\{agent_1,agent_2,…,agent_n \}$, each Agent represented as a quadruple $\{id,position,capacity,attributes\}$, where $id$ is the unique identifier of $agent_i$, $position$ is the initial position of the agent, capacity represents the cargo carrying capacity and type of the agent, and attributes is the other attributes of the agent. In this study, attributes represents the speed of the agent, denoted as velocity. Tasks set are composed of $m$ tasks $\{task_1,task_2,task_3,…,task_m \}$, where each represented as a septuple $\left\{id,position_{start},position_{end},time_{start},time_{end},request \right\}$. Each task is uniquely identified by its id. The task has a pickup location $position_{start}$ and a delivery location $position_{end}$, as well as an arrival time $time_{start}$ and a latest delivery time $time_{end}$. Each task has a cargo type and a requirement request. An agent is eligible to accept the task only if the cargo type matches its own and the remaining capacity is not less than the task’s requirement.
This paper uses a binary variable $x_{ij}$ to represent assigning $task_j$ to $agent_i$, $S_{ij}(\cdot )$to denote the cost $task_j$ of $agent_i$ completing tasks in some order, andwhere the agent and tasks are located $p_i$ indicates the sequence in which $agent_i$ executes tasks. Considering the time effect of completing tasks, the objective function of task allocation is expressed as:
\begin{equation}
    max \sum\nolimits_{i}^{n}\sum\nolimits_{j}^{m} S_{ij} (p_i)x_{ij} \quad s.t. \label{eq1}
\end{equation}
\begin{equation}
    \forall i \in [0,n)   \quad   \sum\nolimits_{j}^{m} request_j x_{ij}  \leq capacity_i \label{eq2}
\end{equation}
\begin{equation}
    \forall i \in [0,n) \quad \forall j \in [0,m) \quad x_{ij} \in \{0,1 \}
\end{equation}
\begin{equation}
    \forall i \in [0,n),j\in [0,m)  start_j \leq t_{ij} (0)<t_{ij} (1) \leq end_j \label{eq4}
\end{equation}
Equation \ref{eq1} formulates the objective function of task assignment as maximizing the total reward score for completing all tasks. Equation \ref{eq2} represents the total demand of tasks accepted by $agent_i$ cannot exceed its carrying capacity. Equation \ref{eq4} specifies that $agent_i$ must arrive at the pickup location no earlier than the task start time, complete the task delivery before the deadline, and ensure that the pickup and delivery locations are different. The generalized MAPD consists of heterogeneous task allocation and heterogeneous path planning for multiple agents. Thus, incorporating heterogeneous MAPF constraints into the existing ones forms the complete set of constraints for the generalized MAPD problem. The constraints of the multi-attribute heterogeneous MAPF problem are as follows:
\begin{equation}
    S_{it}=f(path_{it},attribute_{i}) \label{eq5}
\end{equation}
\begin{equation}
   \forall i, \forall j \in [0, n), i \neq j \quad S_{ij} \cap S_{ji} \neq \oslash \label{eq6}
\end{equation}
\begin{equation}
    attribute_i = attribute_{ig} \cup attribute_{im} \label{eq7}
\end{equation}
Equation \ref{eq5}, \ref{eq6}, and \ref{eq7} indicate that there should be no path conflicts between agents at any timestep. Collectively, Equation 2 to 7 constitute the complete constraints set for the generalized MAPD problem.

\subsection{Step 2: Group Consensus-Based Heterogeneous Auction Algorithm}
The group-consensus based heterogeneous auction algorithm comprises four phases: task processing, task packaging (auction), conflict resolution, and task unpacking/sorting. As shown in Fig. 2, the algorithm incorporates a path planning module to collectively address the generalized MAPD problem. During operation, the system first decomposes tasks exceeding a single agent's capacity into manageable units, then groups similar and proximate tasks according to predefined rules. Each agent bids on task groups, temporarily stores the highest-bid task in its local queue, and ultimately assigns tasks to the highest bidder through consensus-reaching negotiation. Upon consensus, agents receive task clusters requiring decomposition into executable units. Each task contains two ordered waypoints that undergo final sequencing to maximize execution scores.The overall workflow of GCBHA is presented in pseudocode as shown in Algorithm 1(\textnormal{as seen in Appendix}).

\begin{figure*}[!t]
\centerline{\includegraphics[width=0.63\textwidth]{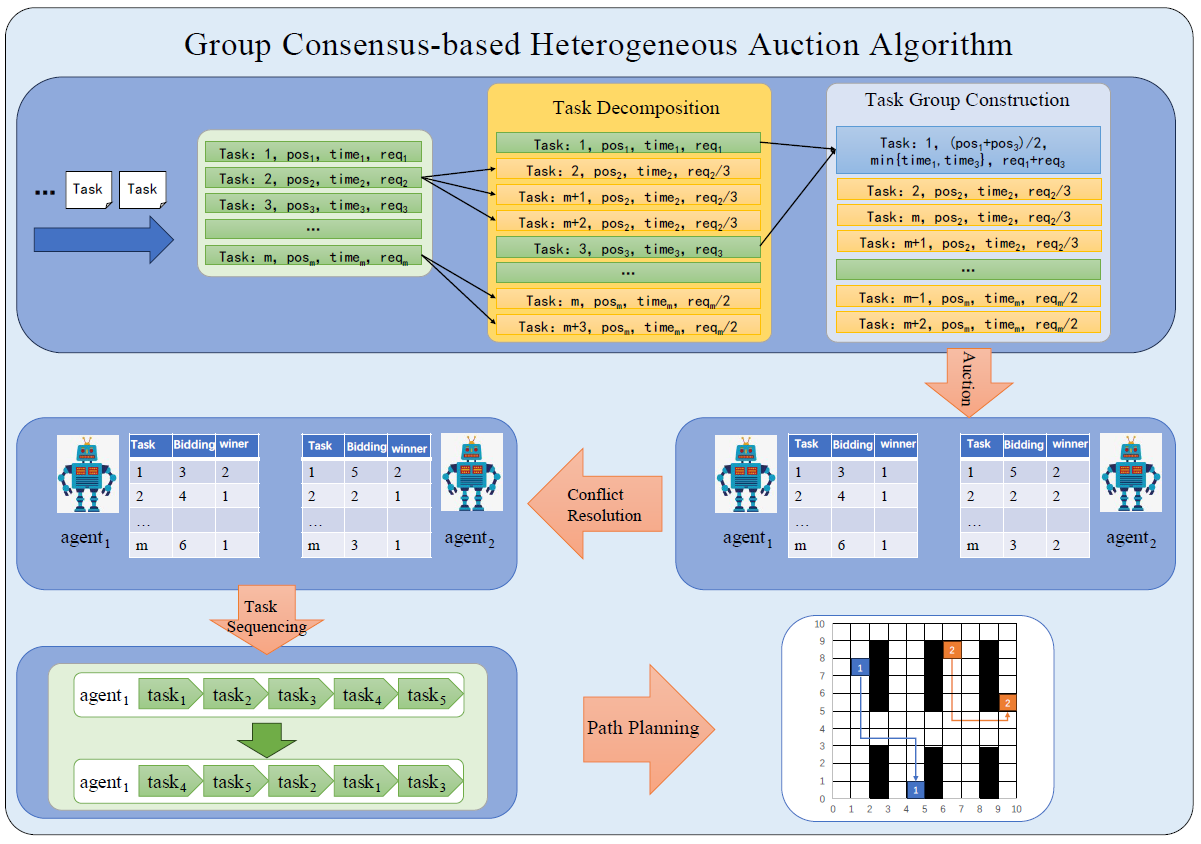}}
\caption{Heterogeneous Auction Based on Group Consensus Algorithm and Path Planning Process.}
\label{fig2}
\end{figure*}

\subsubsection{Task Processing
}

In certain warehouse systems, large-scale order tasks may emerge, making it necessary to break them down into smaller tasks that can be managed by individual agents, as a single agent cannot handle them alone. 
Assuming there is a large task $\left\{id, position_{\text{start}}, position_{\text{end}}, time_{\text{start}}, time_{\text{end}}, request \right\}$, where $request > \max(capacity)$. This task is decomposed into $x$ tasks, which have the same attributes as the original task except for $request$ and $id$. Except for the last task $request_x=request-(x-1)min(capacity)$formed by the decomposition, the rest are task $request_i=min(capacity)$. By decomposing the large-scale task according to the aforementioned rules, we obtain tasks that are identical to the original task except for the task demand and number. Each of these tasks is designed to fit within the carrying capacity of any single agent. Algorithm 2 shows the pseudocode for task decomposition(\textnormal{as seen in Appendix}). 

 In situations where numerous tasks and agents are present, particularly when the task count rises significantly following decomposition, achieving consensus among agents becomes more challenging. Hence, minimizing the number of tasks to expedite consensus is an area for enhancement. This section introduces a task grouping approach grounded in clustering techniques, allocating tasks into groups of a predetermined size.The methodology is detailed in Algorithm 3 (\textnormal{as seen in Appendix}).

\subsubsection{Task Package Construction}
Task packaging constitutes an auction process where agents asynchronously bid on eligible tasks. Each agent maintains two vectors $x_i$ and $y_i$, both of length $m$, where $m$ represents the maximum number of tasks that an agent can accept simultaneously. In this study, $m$ is set to the total number of tasks, as each agent's maximum concurrent task capacity is constrained by its carrying capability.$x_i$ records the current task list of $agent_i$. If the j-th task is assigned to $agent_i$, then $x_{ij}=1$; otherwise,$x_{ij}=0$.$y_i$ records the highest bid value known for each task by $agent_i$.Both vectors are initialized with zero values. Let $c_{ij}$ be the path cost for $agent_i$ to complete the j-th task.

In most studies, $c_{ij}$ is the Euclidean distance from $agent_i$ to $task_j$ \cite{liu2019task}.However, this estimation method may lead to substantial cost prediction errors. To address this limitation, we propose a novel computation approach based on shelf distribution patterns:

\begin{strip}
\begin{equation}
cost(a,b) = 
\begin{cases}
|a.x - b.x| + |a.y - b.y|, & \text{if } |a.x - b.x| > l \\
                          & \text{or } nh < a.y, b.y < (n+1)h \\
                          & \text{or } nw < a.x < (n+1)w \\
                          & \text{or } nw < b.x < (n+1)w \\
|a.y - b.y| + \min(|nw - a.x| + |nw - b.x|), & \text{otherwise}
\end{cases}\label{eq8}
\end{equation}
\end{strip}

% \begin{strip}
% \begin{equation}
% cost(a,b) = 
% % \begin{cases}
% % |a.x - b.x| + |a.y - b.y|, & \text{if } |a.x - b.x| > l \quad \text{or } nh < a.y, b.y < (n+1)h\\
% %                           & \text{or } nw < a.x < (n+1)w 
% %                           \quad \text{or } nw < b.x < (n+1)w \\
% % |a.y - b.y| + \min(|nw - a.x| + |nw - b.x|), & \text{otherwise}
% % \end{cases}\label{eq8}
% \end{equation}
% \end{strip}

\begin{figure}[htbp]
	\centering
	\includegraphics[width=0.85\linewidth]{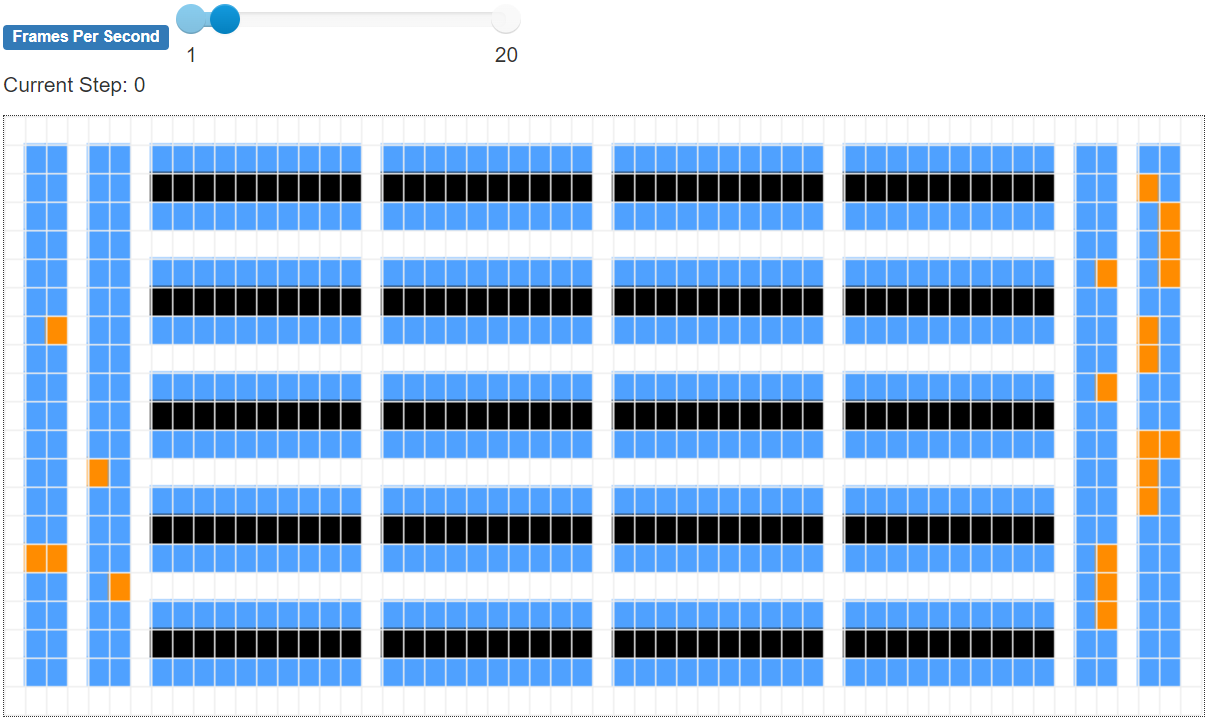}
	\caption{Simulation Diagram of the Warehouse System.}
	\label{fig3}
\end{figure}

Equation \ref{eq8} assumes that the shelves are positioned along the x-axis, as illustrated in Fig. \ref{fig3} where the middle four columns are the shelves. If the goods are arranged along the y-axis, it is only necessary to swap the positions of $x$ and $y$ in the equation.$n$ is a non-negative integer that represents a specific row or column of shelves within the scenario. $h$ represents the y-axis distance between shelves. w represents the x-axis distance between shelves, and $l$ is the length of the shelves. The first equation in Equation \ref{eq8} includes four conditions: 1) The x-coordinates of both points are greater than $l$, indicating that there is a vertical gap between the corresponding shelves. 2) The two points are located within the same horizontal shelf gap. 3) Point a is located within a vertical shelf gap. 4) Point b is located within a vertical shelf gap. These conditions guarantee a translatable rectangular path between points, enabling direct distance calculation via coordinate difference absolutes. Equation 8's second equality handles coordinates in the same shelf column but distinct gaps, where the minimum point distance equals the sum of each point's shortest distance to adjacent vertical gaps.Equation \ref{eq8} is also applied in $nearestTask()$ of Algorithm 2.
In this algorithm, a value attribute value is set for each task, which is directly proportional to the task demand. Let $S_{ij}^n $be the score for $agent_i$ to complete the j-th task in a certain order of task execution.

% \begin{strip}
% \begin{align}
% S_{ij}^n =\ & value_j \times 
% \Bigg\{ 
% e^{-\lambda \left( 
%     \frac{
%         cost(position_i, pn_{start,i}) + time_n - time_{start,i}
%     }{velocity_i} 
% \right)} \notag \\
% &  \ln \Bigg[
% \lambda \left(
%     time_{end,i} -  
%     \frac{
%         cost(position_{start,i}, position_{end,j}) 
%         + cost(position_j, position_{start,i})
%     }{velocity_i}
%     - time_n
% \right) + 1 
% \Bigg]
% \Bigg\}
% \end{align}\label{eq9}
% \end{strip}

\begin{align}
S_{ij}^n & = \text{value}_j \times \Bigg\{ 
e^{-\lambda \left( 
    \frac{
        \text{cost}(\text{position}_i, \text{pn}_{\text{start},i}) 
        + \text{time}_n 
        - \text{time}_{\text{start},i}
    }{\text{velocity}_i} 
\right)} \notag \\
&  \ln \Bigg[
\lambda \Bigg(
    \text{time}_{\text{end},i} 
    - \frac{
        \text{cost}(\text{position}_{\text{start},i}, \text{position}_{\text{end},j}) 
    }{\text{velocity}_i} \notag \\
&  
    - \frac{
        \text{cost}(\text{position}_j, \text{position}_{\text{start},i})
    }{\text{velocity}_i} 
    - \text{time}_n
\Bigg) + 1 \Bigg]
\Bigg\}
\label{eq9}
\end{align}
Where $\lambda \in (0,1]$, and $\text{time}_n$ denotes the completion time of the preceding task. Equation 9 divides the score for completing task $j$ into two parts: the first represents the reward for $\text{agent}_i$ reaching the starting position of the task, and the second corresponds to the reward for $\text{agent}_i$ reaching the end position of the task. The value of $S_{ij}^n$ increases as $\text{agent}_i$ achieves shorter arrival times at both the target's starting point and ending point. $S_{ij}^n$ can be described as the score obtained by inserting the $j$-th task into the $n$-th position in the task queue of $\text{agent}_i$. $S_{ij}$ represents the score $S_{ij}^n$ obtained by inserting the $j$-th task into the task queue in a way that maximizes the total score $S_i$, while keeping the current task queue unchanged. Therefore, if the current task queue of $\text{agent}_i$ remains fixed, $S_{ij}$ is a determined and constant value. $\text{agent}_i$ will only bid for tasks where $S_{ij} > y_{ij}$ and the task can increase the total score $S_i$ of $\text{agent}_i$. Since $s_{ij} \leq 0$, $\text{agent}_i$ will always bid for tasks when it can accept them. The bid value $\text{agent}_i$ for task $j$ is determined by the position where task $j$ can maximize the total score $S_i$ in the task queue of $\text{agent}_i$. After $\text{agent}_i$ bids on all tasks that meet the bidding conditions, $\text{agent}_i$ adds the won tasks into the task queue, thereby completing the task package construction process.

\subsubsection{Conflict Resolution} 
In the previous section, agents did not communicate with each other during the bidding process. This lack of coordination could result in multiple agents simultaneously adding the same task to their individual task queues. In this section, agents communicate to resolve conflicts, ultimately converging on a list of successful bids that determines the winning agent for each task.

Let $G(\tau)$ be a symmetric adjacency matrix representing the undirected communication network between agents at time $\tau$. At time $\tau$, the communication connection between $\text{agent}_i$ and $\text{agent}_j$ is represented by $g_{ij}(\tau)$, where $g_{ij}(\tau) = 1$ if a connection exists between $\text{agent}_i$ and $\text{agent}_j$, and $g_{ij}(\tau) = 0$ otherwise. $\text{agent}_i$ and $\text{agent}_j$ are neighbors if $g_{ij}(\tau) = 1$. Based on the above conditions, it is evident that $g_{ii}(\tau) = 1 \ \forall i$.

When $g_{ij}(\tau) = 1$, $\text{agent}_j$ receives vector $y_j$ from $\text{agent}_i$. Based on all received $y$ vectors, $\text{agent}_j$ releases tasks that have higher bids than its own or replaces tasks that have lower bids. This method is specifically designed for single-task, single-robot scenarios, where each robot can handle only one task at a time. When other agents submit higher bids for task $j$ than $\text{agent}_i$, $\text{agent}_i$ releases $\text{task}_j$ from its task queue. Since task scores are calculated based on estimated arrival times, removing $\text{task}_j$ invalidates the scores of all subsequent tasks in the queue. $\text{agent}_i$ must therefore also release all tasks following $\text{task}_j$ to prevent making decisions based on outdated score information.
However, indiscriminately releasing all subsequent tasks in every case would unnecessarily increase algorithmic complexity. To address this limitation, we modify the consensus strategy. This study adopts the complete consensus strategy provided by CBGA, as detailed in Table \ref{tab1}~\cite{hunt2014consensus} (\textnormal{as seen in Appendix}). 

The CBGA consensus strategy utilizes three key vectors for implementation: the successful bid vector $y_i$, the successful agent vector $z_i$, and the latest time vector $t_i$, which $t_i$ captures the timestamp of the most recent information update from other agents. Three actions are defined:

\begin{itemize}
    \item[1)] \textbf{update:} \( y_{ij} = y_{kj},\ z_{ij} = z_{kj} \)
    \item[2)] \textbf{reset:} \( y_{ij} = 0,\ z_{ij} = 0 \)
    \item[3)] \textbf{leave:} \( y_{ij} = y_{ij},\ z_{ij} = z_{ij} \)
\end{itemize}

The pseudocode for constructing and resolving conflicts in the $t$-th task package are presented in Algorithm 4 (\textnormal{as seen in Appendix}) . 

\subsubsection{Task Unpacking and Sorting}

Once the agents have reached a consensus, they are assigned the tasks they have successfully bid for. However, these tasks are composed of multiple virtual tasks rather than a single executable task, so they need to be dismantled into actual tasks. In Task Grouping, Algorithm 2 returns both a new task list and a grouping list groupList.A preserves multiple tasks that constitute each task group. Based on the task group number id and groupList obtained from the agent’s bid, the actual task list groupList[id] can be obtained. Then, the task groups are removed from each agent’s task queue and the corresponding actual tasks are added. Nevertheless, the tasks resulting from the dismantling process are unsorted and not arranged in the most efficient sequence for execution. Therefore, the system must reorganize each agent's task package to optimize the execution order.

The task sorting process follows a similar approach to task package construction, but differs primarily in its focus on sorting tasks by their target locations. Let $b_i = \{\text{task}_1, \text{task}_2, \dots, \text{task}_n\}$ represent the actual sequence of tasks for $\text{agent}_i$ after the decomposition of tasks, where each task comprises a starting and an ending location. $b_i$ is actually $\{\text{start}_1, \text{end}_1, \text{start}_2, \text{end}_2, \dots, \text{start}_n, \text{end}_n\}$, where $\text{start}_j$ and $\text{end}_j$ are the start and end positions of $\text{task}_j$, respectively, and still retain the time attribute of $\text{task}_j$. The task ordering is essentially the sorting of task positions, with the objective of maximizing the score $S_i$ achieved by $\text{agent}_i$ upon completing all tasks, namely:

\begin{equation}
S_i = \max \left( \sum\nolimits_{j=1}^{n} S_{ij}^{p_i} \right), \quad j \in [1, n)
\label{eq10}
\end{equation}

\begin{equation}
S_{ij}^{p_i} = \text{value}_j \cdot e^{ -\lambda \left( \frac{ \text{cost}(\text{position}_i, \text{target}_j) }{ \text{velocity}_i } + \text{time}_{p_i} \right) }, \quad \text{target}_j \in b_i
\label{eq11}
\end{equation}

Equation \ref{eq11} defines the score for reaching position $\text{target}_j$ in the order of $p_i$. $\text{time}_{p_i}$ denotes the time at which the previous task was completed by $p_i$ in that order. Here, there is a constraint that in $p_i$, $\text{start}_j$ cannot be placed after $\text{end}_j$, because goods can only be delivered after they are picked up.

% Let $p_i$ be the ordered sequence of target positions for $\text{agent}_i$, with $\text{agent}_i$ reaching each target position in the order specified by $p_i$. In each iteration, the algorithm calculates scores for all remaining positions and selects the one that maximizes the total cumulative score. The target position yielding the highest score is inserted into sequence $p_i$ at its optimal position. This insertion completes the task ordering process, thereby concluding the task allocation phase. The complete procedure for task disassembly and sorting is formally described in Algorithm 4.

Let $p_i$ denote the ordered sequence of target positions for $\text{agent}_i$, visited in order. In each iteration, the algorithm scores remaining positions and inserts the one with the highest cumulative score into its optimal place in $p_i$. This process completes task ordering and concludes the allocation phase, as detailed in Algorithm 5.

\subsection{Step 3: Generalized Multi-Agent Pickup and Delivery}

This study proposes a comprehensive GMAPD solution, which integrates a heterogeneous multi-robot task allocation algorithm with a multi-attribute heterogeneous path planning algorithm to achieve end-to-end optimization from task allocation to path planning. During the task allocation phase, the system generates an ordered target location queue for each agent. In the path planning phase, the solution overcomes the limitations of traditional algorithms in handling multi-attribute heterogeneity, specifically addressing the multi-attribute heterogeneous multi-agent path finding (MAH-MAPF) problem.
In the heterogeneous multi-agent path planning problem, agents possess not only traditional kinematic attributes but also geometric properties (e.g., shape, size) and movement speed characteristics. To address the extended MAH-MAPF problem, we enhance the classical conflict-based search (CBS) framework by developing the disjoint splitting forecast conflicts heterogeneous conflict-based search (DSFC-HCBS) algorithm. This approach introduces two key innovations:1) A conflict prediction mechanism extending single-point to dimension-aware area constraints, significantly reducing redundant replanning;2) A disjoint splitting strategy prioritizing positive constraints for agents with larger volumes, higher speeds, or target proximity.These optimizations preserve solution quality while substantially lowering computational complexity, making the DSFC-HCBS algorithm (Algorithm 4) particularly effective for large-scale path planning in practical applications.
Our GMAPD problem incorporates a lifelong MAPF scenario where agents continuously receive new targets upon reaching their destinations. The primary challenge in this framework arises from agents' asynchronous arrival times at their targets, which necessitates immediate path replanning to subsequent destinations - the core difficulty of lifelong MAPF problems. This characteristic requires specific adaptations to our DSFC-HCBS algorithm.
After task assignment, each agent maintains an ordered target queue, with the final step involving collision-free path planning for all agents using our hierarchical DSFC-HCBS method\cite{jiang2024improve}.Unlike traditional one-time MAPF solutions, this lifelong paradigm introduces the key challenge of asynchronous goal completion. When an agent reaches its target, the system must instantly compute its path to the next destination - the central computational demand of lifelong MAPF. To meet these requirements, we have modified the DSFC-HCBS algorithm accordingly.

Initially, all agents execute standard one-time MAPF from designated starting positions. When some agents reach goals early, naive full replanning may require up to $m-n$ iterations for $n$ agents and $m$ tasks. Given the high computational cost of path planning, especially in complex environments, reducing redundant replanning is essential.

% In the initial state, all agents start at their designated positions, following standard one-time MAPF procedures for the first path planning iteration. However, when certain agents reach their goals before others during this initial phase, a naive approach would involve replanning paths for all agents. While feasible, this strategy results in significant computational redundancy---potentially requiring up to $m-n$ planning iterations for $n$ agents and $m$ tasks. Given the substantial computational overhead of path planning, particularly in complex environments, minimizing redundant planning operations becomes crucial.

When an agent reaches its target location, the system automatically initiates a new path planning cycle. In this process:1) The paths that were planned previously but not yet implemented are incorporated as constraints into the DSFC-HCBS constraint set. 2) New routes are computed for all agents to reach their updated objectives. 
This approach leverages the fact that existing planned paths already represent valid, collision-free solutions to agents' current destinations. However, since some agents have not yet arrived at their previous destinations and must continue along their predetermined paths, a time delay $\tau$ is necessary when plotting their courses to the new targets. $\tau$ represents the duration these agents need to reach their current objectives from their present positions.

% The primary advantage of this methodology lies in its decoupling of path planning complexity from system scale---planning frequency depends exclusively on an agent's maximum task capacity rather than the total number of agents or tasks. This approach drastically reduces the number of path planning iterations and shortens the algorithm’s execution time. The integration of GCBHA and the lifelong DSFC-HCBS framework thus provides an effective solution to the GMAPD challenge.

This method's core advantage lies in decoupling path planning complexity from system scale—planning frequency depends solely on agent task capacity, not total agents or tasks. This significantly reduces planning iterations and execution time. Integrating GCBHA with the lifelong DSFC-HCBS framework offers an effective solution to the GMAPD problem.

% \begin{table}[htbp]
%     \centering
%     \caption{}
%     \label{tab2}
%     \begin{tabular}{p{2.8cm} p{4.2cm} p{7.2cm}}
%         \toprule
%         \textbf{Experiment Number} & \textbf{Experiment Objective} & \textbf{Experiment Method} \\
%         \midrule
%         Experiment 1 &
%         Test the performance of GCBHA in solving task allocation problems. &
%         Compare the algorithm runtime and the total score of the solutions required for task allocation by GCBHA, CBGA, and nCAR under the conditions of varying numbers of tasks and agents, as well as whether the carrying capacity of the agents is considered. \\
        
%         Experiment 2 &
%         Evaluate the performance of GCBHA in resolving GMAPD issues, and verify the precision of the distance estimation method in warehouse scenarios. &
%         Compare the performance of CENTRAL, CBGA, GCBHA, and TA-priority in solving GMAPD under different parameter conditions, and summarize the algorithm runtime, the length of the solution path, as well as the difference between the estimated path length and the actual path length. \\
%         \bottomrule
%     \end{tabular}
% \end{table}

\subsection{Step4: Construction of the Experimental System}
Traditional task allocation algorithms typically evaluate performance through predicted cost comparisons without simulating physical agent movement. In contrast, our computational experiments establish a warehouse simulation environment where agents dynamically accept tasks and navigate freely. This allows researchers to investigate algorithm performance across a range of scenarios and provides a clear comparison of the actual time or path cost for robots to complete tasks. Figure 2 shows our warehouse simulation environment, designed to match the research context. In the scene, the optional task endpoints are located on both sides, the middle black area represents the shelves, and the pickup points for tasks are situated alongside them. The start and end positions of tasks, as well as the initial position of the agent, are all randomly selected from the set of available positions. After task allocation, the agent moves along the planned path at its own speed. The specific tasks and agent structures are identical to the definitions of tasks and agents in Step 1. At initialization, a specified number of tasks and agents are randomly generated by the system or read from a file. Subsequently, the task allocation algorithm is employed to distribute tasks. Once task allocation is completed, the DSFC-HCBS algorithm is used for path planning. Finally, the agent moves to the target location according to the planned path to execute the task.

To evaluate GCBHA performance, comprehensive experiments were conducted in two phases:1) Comparative analysis against state-of-the-art task allocation algorithms;2) Integration with MAH-MAPF forming a complete GMAPD solution, benchmarked against existing MAPD methods. Given task allocation's critical role in MAPD, algorithms are distinguished by their allocation methods. Two sets of experiments were designed. Experiment 1: To evaluate the performance of GCBHA in solving task allocation problems. Experiment 2: To assess the performance of GCBHA in addressing the complete GMAPD problem. The specific experimental designs and parameters are shown in Table\ref{tab2} and Table\ref{tab3} (\textnormal{as seen in Appendix}). 
%All algorithms in this study were implemented in Python and tested on a laptop with 16GB of RAM and a 2.30 GHz i7-12700H CPU.

\section{EXPERIMENT RESULTS AND ANALYSIS}

\subsection{Experiment 1: Task Allocation Performance of GCBHA}

\begin{figure}
    \centering
    \includegraphics[width=0.8\linewidth]{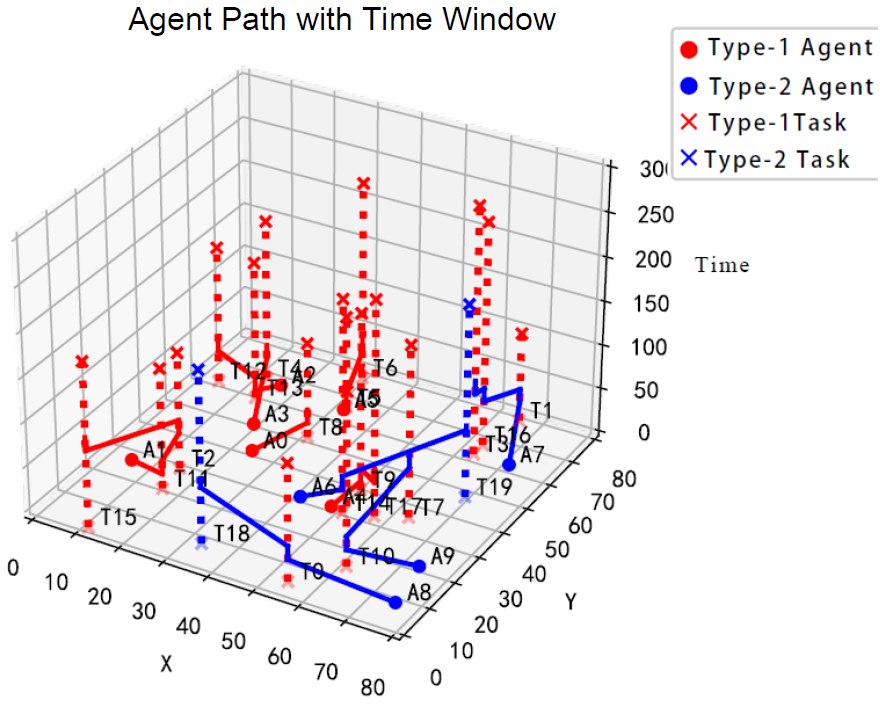}
    \caption{Agent Routes in CBGA.}
    \label{fig4}
\end{figure}

\begin{figure}
    \centering
    \includegraphics[width=0.8\linewidth]{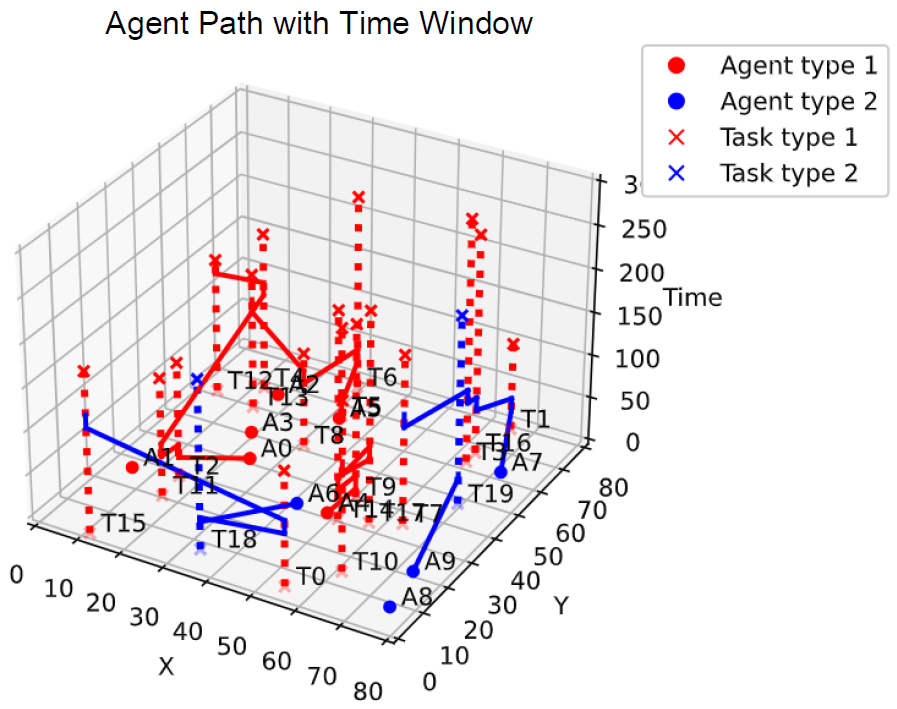}
    \caption{Agent Scheduling Time in CBGA.}
    \label{fig5}
\end{figure}

\begin{figure}
    \centering
    \includegraphics[width=0.8\linewidth]{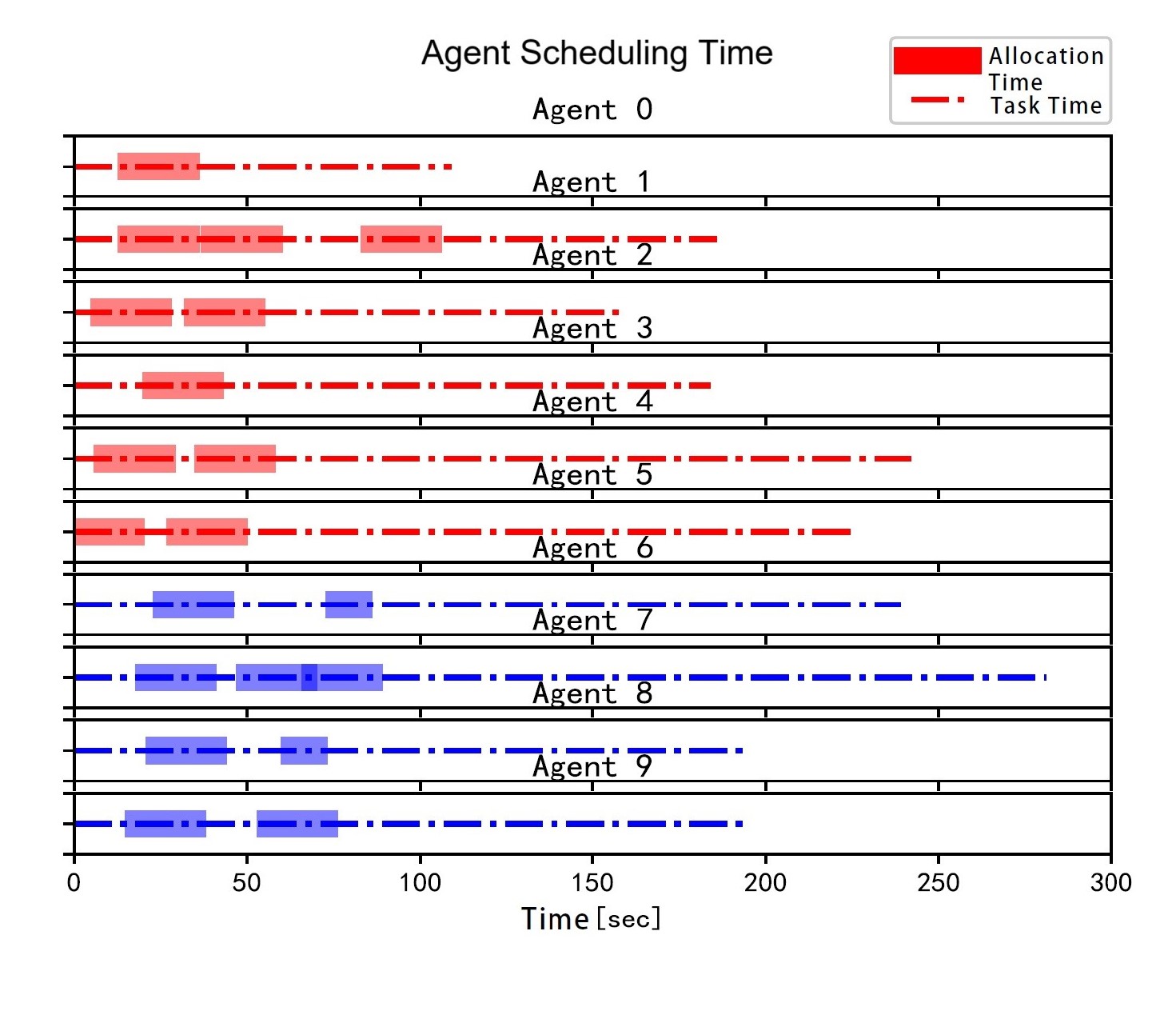}
    \caption{Agent Scheduling Time in CBGA.}
    \label{fig6}
\end{figure}

\begin{figure}
    \centering
    \includegraphics[width=0.8\linewidth]{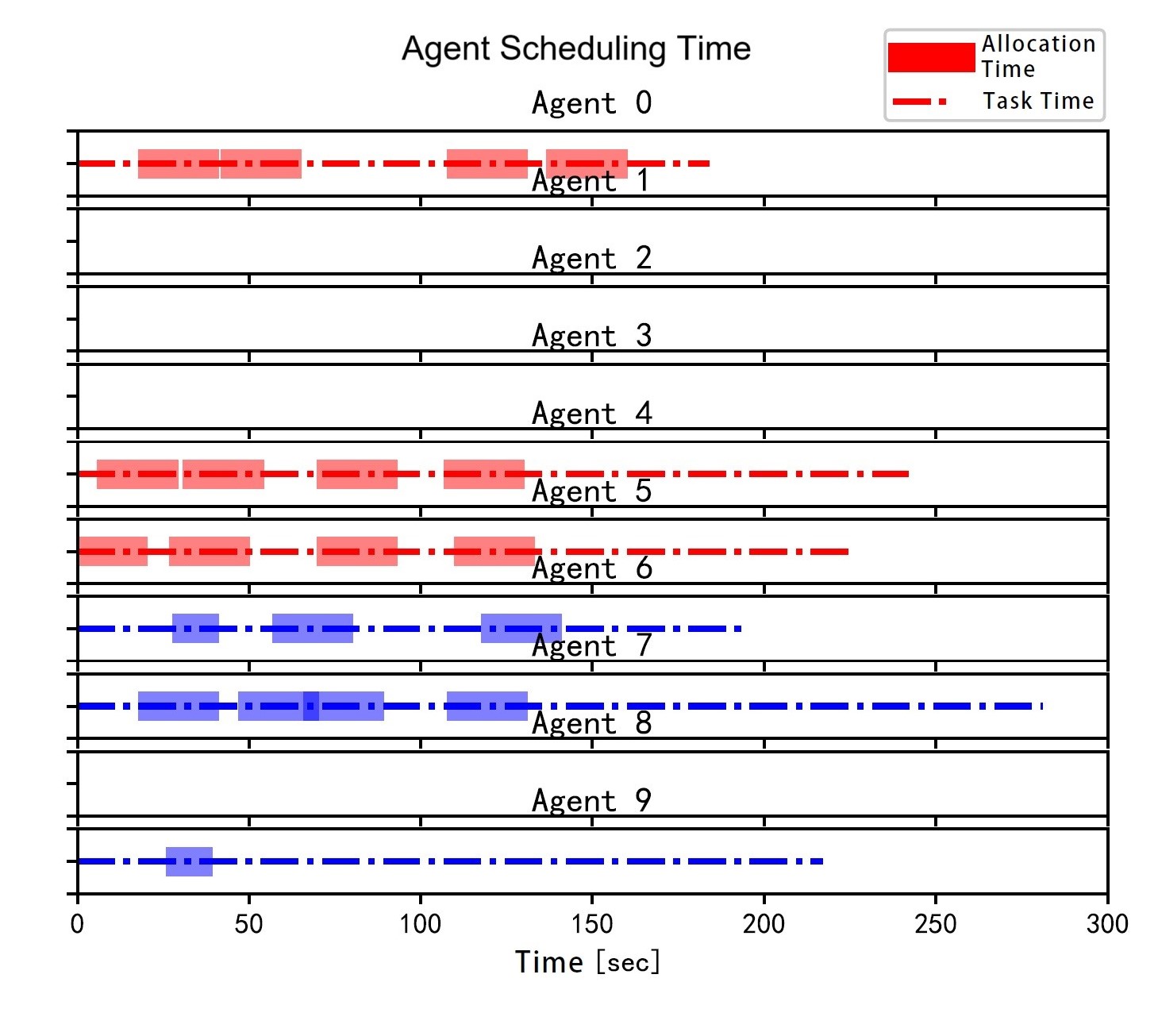}
    \caption{Agent Scheduling Time in GCBHA.}
    \label{fig7}
\end{figure}

Fig.\ref{fig4} and \ref{fig5} respectively present the route maps of agents after task allocation under the conditions of 10 agents and 20 tasks, using CBGA and GCBHA with a group task requirement of 50. The route maps visualize both the arrival time at each target location and the spatial positions of targets. Fig. \ref{fig6} and \ref{fig7} present the task execution times for all agents. As can be clearly observed from Fig. \ref{fig7}, some agents are not assigned tasks, and the two algorithms exhibit fundamentally different task allocation patterns. There is a situation where agents are not assigned tasks by GCBHA because the number of tasks after grouping is not enough to be assigned to each agent. From the above results, it can be concluded that when the number of tasks is too small, an excessively high group task requirement negatively impacts allocation outcomes, and the group task requirement should be appropriately reduced. However, in practice, the path lengths of GCBHA solutions do not differ significantly from those of CBGA under such conditions, because the tasks in each group are the ones with the lowest cost. Nonetheless, there is a certain increase in task completion time.

\begin{figure*}
    
    \includegraphics[width=1\linewidth]{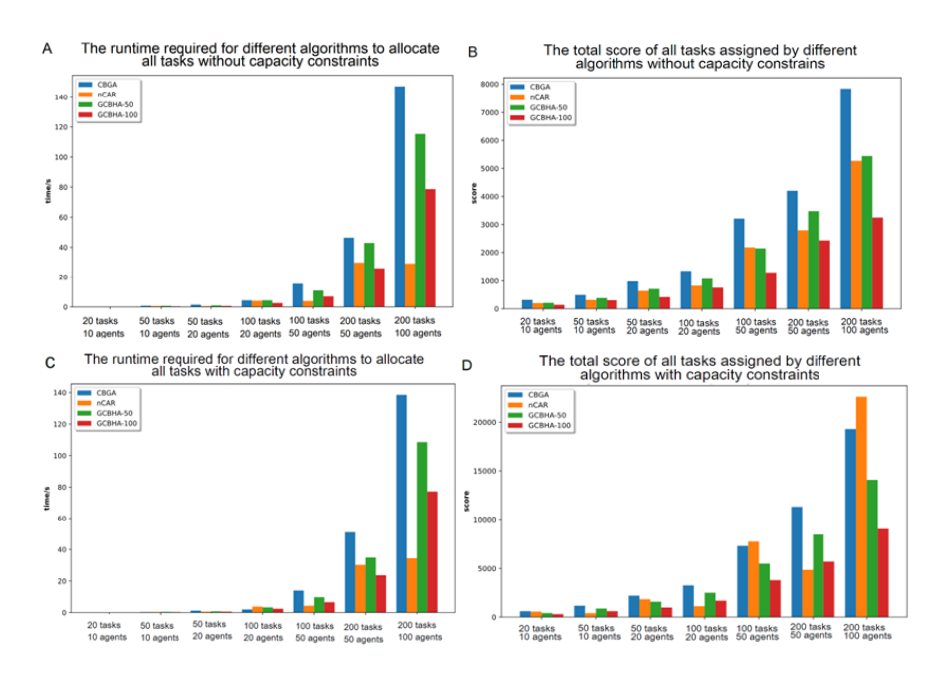}
 \captionsetup{justification=justified, singlelinecheck=false}
    \caption{The runtime and the total score of all tasks assigned by different algorithms. Fig.8A: The runtime of different algorithms for completing all task assignments without considering agent carrying capacity. Fig. 8B: Total score of different algorithms for completing all task assignments without considering agent carrying capacity. Fig. 8C: The runtime required for different algorithms for completing all task assignments when considering agent carrying capacity. Fig. 8D: Total score of different algorithms for completing all task assignments when considering agent carrying capacity.}
    \label{fig8} 
\end{figure*}

Fig. 8(A) and 8(C) present the runtime and total solution scores of the GCBHA, CBGA, and nCAR\cite{sarkar2018scalable} algorithms under varying parameter conditions when agent carrying capacity and task requirements are disregarded. With capacity constraints removed, all tasks were uniformly valued at 100 for standardized comparison. The figures demonstrate the performance of the GCBHA algorithm under different group task requirements. Fig. 8(A) and 8(C) demonstrate that increasing the group task requirement in GCBHA leads to a trade-off between computational efficiency and the total solution score. Compared to CBGA, task grouping inevitably reduces the solution score because agents are no longer completely free to accept tasks. GCBHA trades a slight degradation in solution quality for improved runtime speed. In practical, optimal solutions are often unnecessary, and algorithms capable of quickly computing suboptimal solutions are more practically valuable. nCAR performs well in both runtime and solution quality since it is a centralized algorithm that eliminates agent communication and consensus steps, significantly reducing computational overhead. However, nCAR struggles to scale to large scenarios and relies on robust communication infrastructure. In contrast, GCBHA imposes lower demands on communication quality and requires no synchronous iterations among agents.
Fig. 8(B) and 8(D) illustrate the solution time and total score of solutions for the GCBHA, CBGA, and nCAR algorithms under different parameter conditions when considering agent carrying capacity and task requirements. Compared to Fig. 8, it can be concluded that considering agent carrying capacity has almost no impact on algorithm solution time. The solution time remains negatively correlated with group task requirements, while the total score shows a positive correlation with these requirements.  The total score of nCAR solutions exhibits significant sensitivity to both the number of tasks and agents, which is attributable to task locations. By prioritizing assigning agents to their nearest tasks, nCAR achieves maximum scores for individual task completions, thus enhancing the overall total score.

The experimental results demonstrate that GCBHA achieves an optimal balance between computational efficiency and solution quality, with minimal degradation from task grouping. The algorithm maintains solution integrity while allowing performance to be tuned via group task requirements, which can be dynamically adjusted based on the number of tasks and agents. This ensures sufficient tasks per agent post-grouping and helps reduce overall completion time.

% The experimental results demonstrate that GCBHA effectively balances computational efficiency and solution quality in task allocation, without a substantial decrease in solution quality due to task grouping. Based on comprehensive experimental results, it can be concluded that GCBHA achieves an optimal balance between solution efficiency and quality in task allocation problems. The algorithm maintains solution integrity without incurring significant degradation caused by task grouping, while its performance remains controllable through group task requirements. The group task requirements can be dynamically adjusted based on the actual number of tasks and agents, ensuring that there are sufficient tasks for each agent after grouping, and thus reducing the overall task completion time.

\subsection{ Experiment 2: The Algorithm Performance of GCBHA in Solving GMAPD Problems}

Fig. 9 demonstrates the path lengths required for agents to complete all tasks using different algorithms with 20 and 50 agents. It can be observed that the centralized algorithm has the shortest path length among all algorithms, while GCBHA achieves the shortest path length among distributed algorithms. The path length difference between CBGA and TA-priority is not substantial. Fig. 10 illustrates the difference between the path length required for agents to complete all tasks and the path lengths predicted by various algorithms with 20 and 50 agents. GCBHA exhibits the smallest difference between estimated and actual path lengths, while CBGA and TA-priority show significant differences. These experimental results validate the accuracy of the distance estimation method proposed in this paper for warehouse scenarios. Combined with Fig. 9, it is evident that the estimated path length influences the quality of task allocation solutions to a certain extent. A smaller gap between estimation and the actual value can lead to more precise task allocation and reduces the path length for agents to complete tasks.

\begin{figure}[htbp]
  \centering
  \subfigure[20 agents]{\label{fig9_a}
      \includegraphics[width=0.46\linewidth]{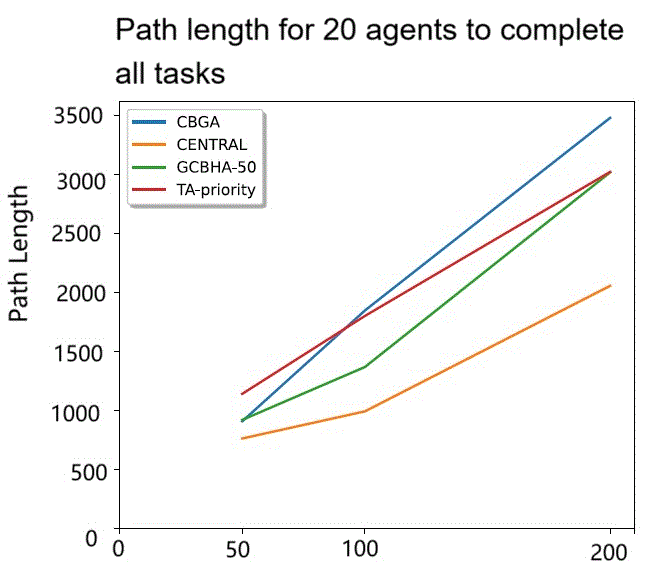}
  }  
  \hspace{0.0\linewidth}
  \subfigure[50 agents]{\label{fig9_b}
      \includegraphics[width=0.46\linewidth]{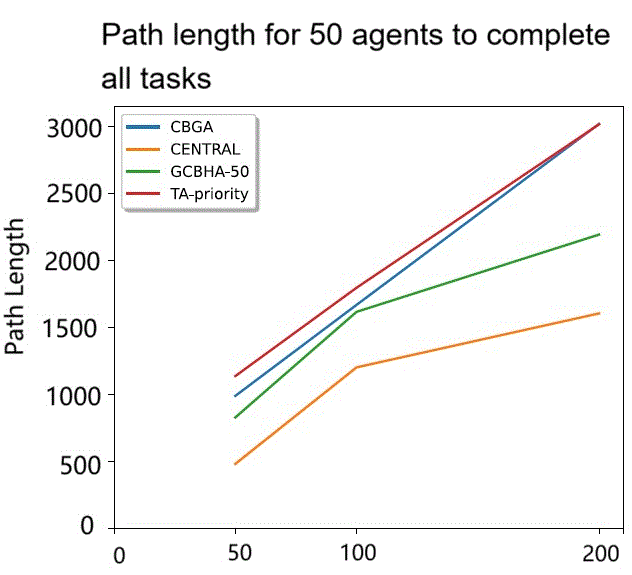}
  }
  
  \caption{Path length required for agents to complete all tasks.}
\end{figure}

\begin{figure}[htbp]
  \centering
  \subfigure[20 agents]{\label{fig10_a}
      \includegraphics[width=0.46\linewidth]{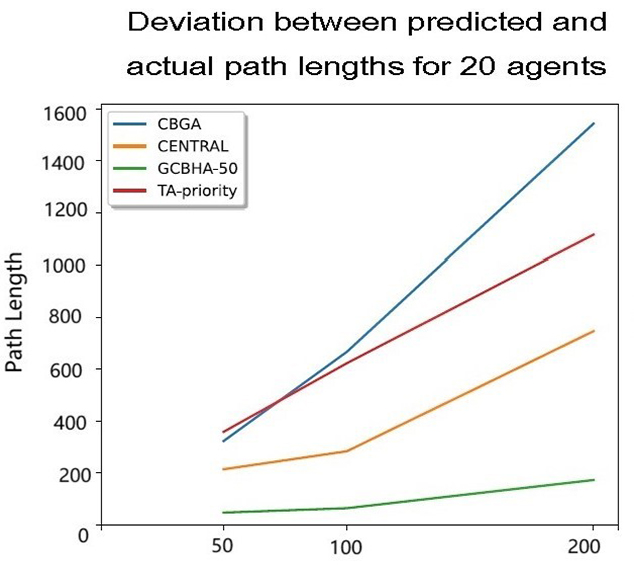}
  }  
  \hspace{0.0\linewidth}
  \subfigure[50 agents]{\label{fig10_b}
      \includegraphics[width=0.46\linewidth]{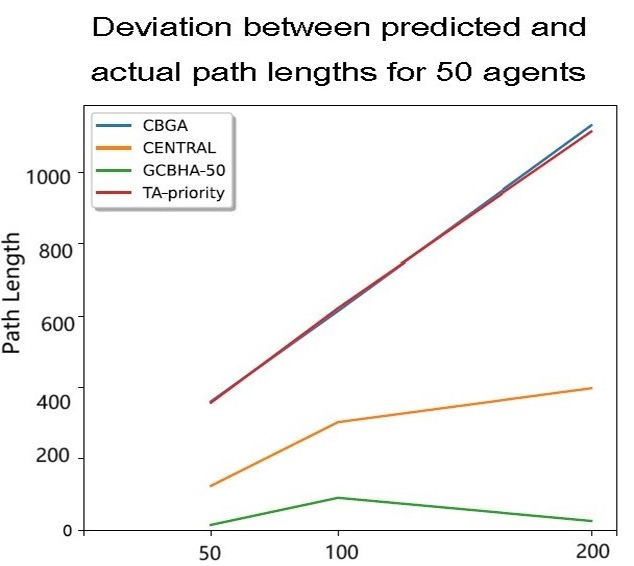}
  }
  
  \captionsetup{justification=justified, singlelinecheck=false}

  \caption{Difference between the actual and predicted path length for agents to complete all tasks}
\end{figure}

\begin{figure}[htbp]
  \centering
  \subfigure[20 agents]{\label{fig11_a}
      \includegraphics[width=0.46\linewidth]{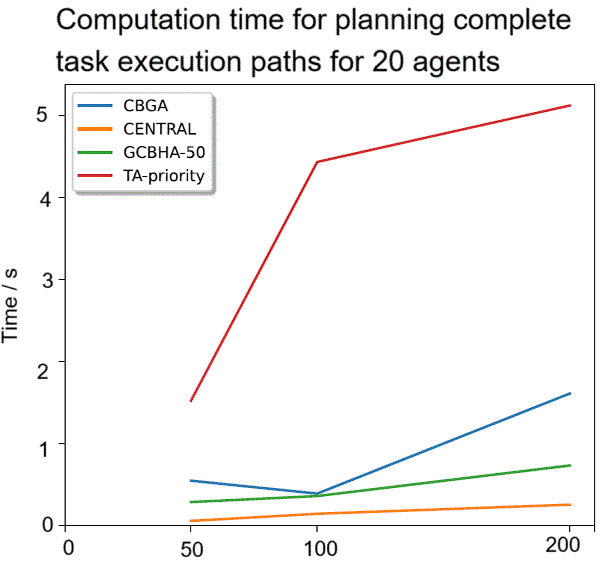}
  }  
  \hspace{0.0\linewidth}
  \subfigure[50 agents]{\label{fig11_b}
      \includegraphics[width=0.46\linewidth]{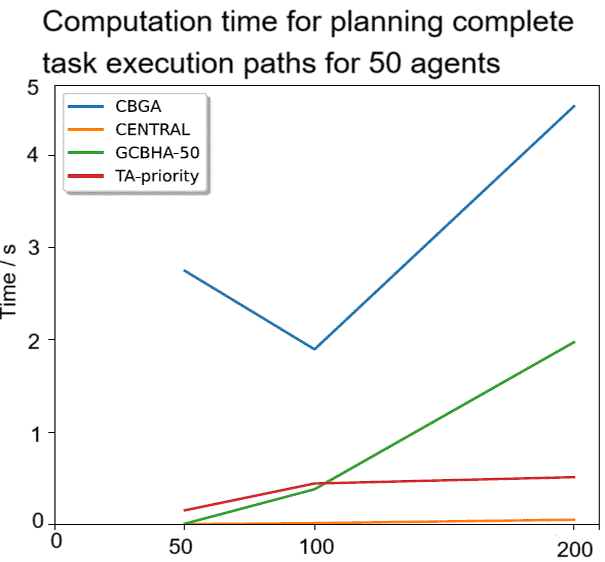}
  }
 
    \captionsetup{justification=justified, singlelinecheck=false}
 
  \caption{Runtime of different algorithms for planning all paths to all target positions.}
\end{figure}

\begin{figure}[htbp]
  \centering
  \subfigure[20 agents]{\label{fig12_a}
      \includegraphics[width=0.46\linewidth]{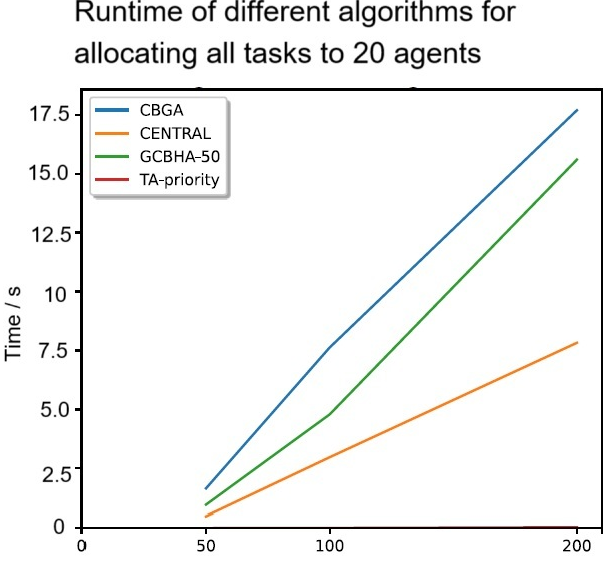}
  }  
  \hspace{0.0\linewidth}
  \subfigure[50 agents]{\label{fig12_b}
      \includegraphics[width=0.46\linewidth]{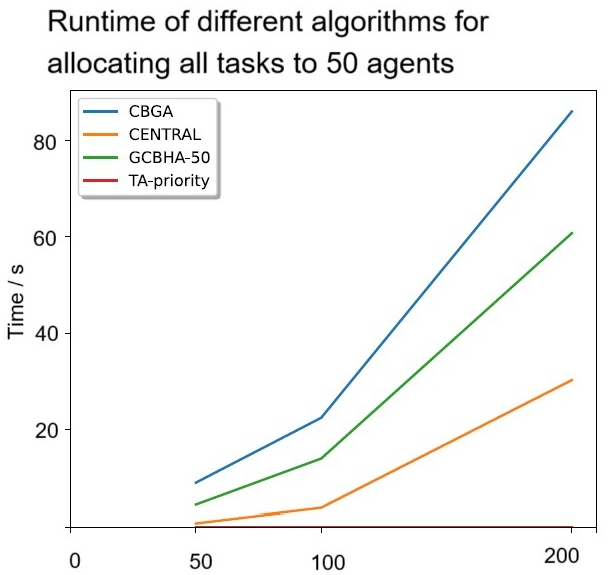}
  }
 
    \captionsetup{justification=justified, singlelinecheck=false}
 
  \caption{Runtime of different algorithms for allocating all tasks.}
\end{figure}
Fig. 11 illustrates the time required by the algorithm to plan paths for all agents after task allocation. Path planning is also a critical element in the MAPD methods, and the complexity of the problem influences the time required for path planning.

Experimental results demonstrate GCBHA's superior efficiency, generating simpler solutions with shorter planning times. In contrast, CBGA and TA-priority exhibit unstable performance. As Fig.12 shows, TA-priority achieves near-instantaneous task allocation through priority-based assignment, but this results in suboptimal allocations requiring longer execution paths.
Distributed algorithms still consume more time for task allocation compared to centralized algorithms. Compared to CBGA, the grouping strategy in GCBHA reduces the time needed to reach consensus. Based on the comprehensive experimental results, GCBHA demonstrates superior performance in both solution quality and algorithm runtime, second only to centralized algorithms.

\section{Conclusion}
Addressing the understudied challenge of heterogeneous multi-task allocation in multi-robot systems, this study proposes a GCBHA algorithm. GCBHA addresses the collaboration challenges of large-scale tasks by task decomposition and employs heuristic clustering grouping to reduce the time required for agents to reach consensus while ensuring solution quality. The proposed method enhances the accuracy of distance cost estimation in task allocation through a scenario-specific distance estimation approach, thereby reducing actual path costs. Experimental results demonstrate that GCBHA achieves superior performance in both solution path length and algorithm runtime for task allocation problems. We believe that this study provides novel insights into the GMAPD heterogeneous multi-task multi-robot task allocation problem and demonstrates potential for addressing practical challenges in item transportation within industrial systems.

%\section{Acknowledgments}
%This work has been supported in part by National Natural Science Foundation of %China (No. 62472306, No. 62441221, No. 62206116), Tianjin University's 2024 %Special Project on Disciplinary Development (No. XKJS-2024-5-9), Tianjin %University Talent Innovation Reward Program for Literature \& Science Graduate %Student (C1-2022-010), and Henan Province Key Research and Development Program %(No.251111210500).

\bibliographystyle{IEEEtran}
\bibliography{IEEEexample}
\clearpage
\appendices
\section*{Appendixes}

\subsubsection{Algorithm 1}
Algorithm 1 illustrates the overall GCBHA procedure in pseudocode.
\begin{algorithm}
	\caption{Consensus-Based Grouping Algorithm}
	\begin{algorithmic}[1]
		\Require $G(V,E)$, $Agents$, $Tasks$  \Comment{Input graph, agents and tasks}
		\Ensure Sorted target position queue  \Comment{Output}
		
		\Statex \textbf{Global:} \textit{TaskAssignment(Tasks, Agents)}
		\Statex
		
		\Function{TaskProcessing}{$Tasks$, $Agents$}
		\For{each $task \in Tasks$}
		\If{$task.request > \max(Agents.capacity)$}
		\State $subtasks \gets task.decompose()$  
		\Comment{Corresponding to Algorithm 2}
		\For{each $subtask \in subtasks$} 
		\State Match $subtask$ to $clusters$  
		\Comment{Using Algorithm 3 rules}
		\EndFor
		\EndIf
		\EndFor
		\State \Return $Tasks$
		\EndFunction
		\Statex
		
		\Function{TaskPackageConstruction}{$Tasks$, $Agents$} 
		\For{each $task \in Tasks$}
		\If{$task.request \leq Agent.capacity$}
		\State $Agent_{i} \gets \text{bid}(task)$
		\If{$Agent_{i}$ wins $task$}
		\State $taskQueue.enqueue(task)$
		\EndIf
		\EndIf
		\EndFor
		\EndFunction
		\Statex
		
		\Function{ConflictResolution}{$Tasks$, $Agents$}
		\If{auction.conflict $==$ True}
		\State initiate\_agent\_communication()
		\EndIf
		\EndFunction
		\Statex
		
		\Function{TaskUnpackingSorting}{$Tasks$, $Agents$}
		\For{each $taskSet \in$ consensus\_allocations}
		\State $subtasks \gets decompose(taskSet)$  
		\State $sortedTargets \gets sort(subtasks)$
		\EndFor
		\State \Return $sortedTargets$
		\EndFunction
	\end{algorithmic}
\end{algorithm}

\subsubsection{Algorithm 2}
Algorithm 2 presents the pseudocode for task decomposition.
\begin{algorithm}
	\caption{Task Decomposition}
	\textbf{Symbol:} \textit{task Decomposition(Tasks, Agents)} \\
	\textbf{Input:} Task list \textit{Tasks}, agent list \textit{Agents} \\
	\textbf{Output:} Task list \textit{Tasks}
	\begin{algorithmic}[1]
		\Function{TaskDecomposition}{Tasks, Agents}
		\For{each $task \in Tasks$}
		\If{$task.request > \max(Agents.capacity)$}
		\State \Comment{If task is not fully decomposed, continue to create newTask}
		\State $newTask \gets task$
		\State $newTask.request \gets \min(Agents.capacity)$
		\State \Comment{The last task requirement is equal to the remaining task requirements}
		\State $newTask.id \gets \text{length}(Tasks)$
		\State $Tasks.\text{add}(newTask)$
		\EndIf
		\EndFor
		\State \Return $Tasks$
		\EndFunction
	\end{algorithmic}
\end{algorithm}

\subsubsection{Algorithm 3}
\begin{algorithm}
	\caption{Task grouping method based on clustering}
	\textbf{Symbol}: \texttt{group(Tasks, request\_group)} \\
	\textbf{Input}: Task list \texttt{Tasks}, Maximum task demand for a group \texttt{request\_group} \\
	\textbf{Output}: Task list \texttt{Tasks}, group list \texttt{groupList}
	\begin{algorithmic}[1]
		\Function{group}{Tasks, request\_group}
		\State \texttt{groupList} $\gets$ \texttt{list()} 
		\State \texttt{copyTasks} $\gets$ \texttt{Tasks}
		\While{\texttt{copyTasks} $\neq$ \texttt{empty}}
		\State \texttt{group} $\gets$ \Call{nearestGroup}{copyTasks, request\_group}
		\State \texttt{groupList.add(group)}
		\State \texttt{copyTasks.remove(group)}
		\EndWhile
		\State \texttt{newTasks} $\gets$ \texttt{list()}
		\For{\texttt{group} in \texttt{groupList}}
		\State \texttt{newtask} $\gets$ \texttt{task()}
		\State \texttt{newtask.request} $\gets$ $\sum$ \texttt{task.request} \textbf{each task} $\in$ \texttt{group}
		\State \texttt{newtask.time} $\gets$ $\min$(\texttt{task.time}) \textbf{each task} $\in$ \texttt{group}
		\State \texttt{newtask.pos} $\gets$ $\operatorname{avg}$(\texttt{task.pos}) \textbf{each task} $\in$ \texttt{group}
		\State \texttt{newTasks.add(newtask)}
		\EndFor
		\State \Return \texttt{newTasks, groupList}
		\EndFunction
		
		\vspace{1em}
		
		\Function{nearestGroup}{Tasks, r}
		\State \texttt{cost\_min} $\gets$ \Call{singleTaskCycle}{Tasks}
		\For{\texttt{each task} $t \in$ \texttt{Tasks}}
		\State $A \gets \{t\}$
		\State $\hat{A} \gets$ \texttt{Tasks.remove(task)}
		\State \texttt{request\_total} $\gets$ \texttt{t.request}
		\While{\texttt{request\_total} $<$ r}
		\State \texttt{newtask} $\gets$ \Call{nearestTask}{$A$, $\hat{A}$, \texttt{r - request\_total}}
		\If{\texttt{newtask = None}} \State \textbf{break} \EndIf
		\State \texttt{request\_total} $\gets$ \texttt{request\_total + newtask.request}
		\State $A.\texttt{add(newtask)}$
		\State $\hat{A}.\texttt{remove(newtask)}$
		\State \texttt{cost} $\gets$ \Call{allTasksCycle}{$A$} + \Call{singleTaskCycle}{$\hat{A}$}
		\If{\texttt{cost < cost\_min}}
		\State \texttt{cost\_min} $\gets$ \texttt{cost}
		\State $A_\text{min} \gets A$
		\State $\hat{A}_\text{min} \gets \hat{A}$
		\EndIf
		\EndWhile
		\EndFor
		\State \Return $A_\text{min}, \hat{A}_\text{min}$
		\EndFunction
	\end{algorithmic}
\end{algorithm}

%In situations where numerous tasks and agents are present, particularly when the task count rises significantly following decomposition, achieving consensus among agents becomes more challenging. Hence, minimizing the number of tasks to expedite consensus is an area for enhancement. This section introduces a task grouping approach grounded in clustering techniques, allocating tasks into groups of a predetermined size. The methodology is detailed in the pseudocode of Algorithm 2. 
The pseudocode for clustering-based task grouping is shown in Algorithm 3.
Given a list of  Tasks and a group task demand size $request_{group}$ ,the total task demand within the group should not exceed $request_{group}$. The value of $request_{group}$ can be determined by the agent based on its carrying capacity and the task demand. A larger value of $request_{group}$ leads to a shorter consensus time for the algorithm but results in a higher path cost.$request_{group}$ should be more than double the minimum task demand and is often set to match the agent’s minimum carrying capacity, ensuring that the agent only needs to accept one set of tasks. The objective of this method is to group tasks according to requirements while minimizing the cost of completing the task groups. Task grouping is based on two main criteria: 1) The total task demand in the group must not exceed the predefined group task demand, and all tasks in the group should be of the same type. 2) When adding tasks to the group, the focus is always on minimizing the cost of completing all tasks within that group.

  In algorithm 3, $nearestGroup()$ provides a method to group tasks with minimal cost. The algorithm initializes with all tasks unassigned. In each iteration: 1) A seed task is assigned to a new target group. 2) Based on this task, searches for the nearest unassigned task of the same type that meets the task demand. This task is added to the group. 3)The process continues until no suitable tasks are found, completing one task grouping. 4)Recalculate the total cost of completing the tasks in the group and the unassigned tasks, denoted as cost, if $cost<cost_{min}$, it proves that this grouping is effective, and the grouping and its cost are recorded. Once all iterations are completed, the set of tasks with the lowest cost will have been identified.

Group tasks using $nearestGroup()$ until the task list is empty. After grouping, add each group of tasks to the task list in the form of task. The newly generated task's demand should be the sum of all individual task demands within the group, while its start time and end time should be set to the earliest start time and the earliest end time among the tasks in the group, respectively. The purpose of this setting is to ensure that when an agent accepts the task group, it can complete all tasks within the allotted time. This method allows the agent to treat the task group as a single task during the auction.
\subsubsection{Algorithm 4}
Algorithm 4 presents the pseudocode for constructing and resolving conflicts in the $t$-th task package. In line~15, \texttt{consensus()} is a function that constructs consensus between $\text{agent}_i$ and $\text{agent}_k$ according to the rules in Table \ref{tab1}, and the return value $J$ represents the modified task number. Lines~16--20 indicate that if there is a modified task in the task package, that task and all subsequent tasks should be removed from the task queue. The algorithm is repeated until all agents reach a consensus on the successful bid vector and the successful agent vector.
\begin{algorithm}
	\caption{\textit{t-th} task package construction and conflict resolution of $agent_t$}
	\textbf{Symbols}: \texttt{bundle(Tasks, $\mathbf{y}_t$, $\mathbf{z}_t$, $b_t$, $\mathbf{y}_k$, $\mathbf{z}_k$)} \\
	\textbf{Input}: Task list \texttt{Tasks}, winning bid vector $\mathbf{y}_t$, winning agent vector $\mathbf{z}_t$, current task queue $b_t$, received winning bid vector $\mathbf{y}_k$ and agent vector $\mathbf{z}_k$ \\
	\textbf{Output}: accepted task queue $b_t$
	\begin{algorithmic}[1]
		\Function{bundle}{$Tasks, \mathbf{y}_t, \mathbf{z}_t, b_t, \mathbf{y}_k, \mathbf{z}_k$}
		\State $\mathbf{y}_t \gets \mathbf{y}_t(t-1)$
		\State $\mathbf{z}_t \gets \mathbf{z}_t(t-1)$
		\State $b_t \gets b_t(t-1)$
		\While{$b_t.request \leq agent_t.capacity$}
		\For{each $j \in \text{length}(Tasks)$}
		\State $c_{ij} \gets s_{ij}(b_t)$
		\State $h_{ij} \gets (c_{ij} > \mathbf{y}_{tj})$
		\State $J_t \gets \texttt{argmax}(c_{ij} h_{ij})$
		\State $n_i \gets \texttt{argmax}(s^2_{ij})$
		\State $b_t.\texttt{add}(Tasks[J_t], n_i)$
		\State $\mathbf{y}_{tj} \gets c_{ij}$
		\State $\mathbf{z}_{tj} \gets i^t$
		\State $b_t.request \gets b_t.request + Tasks[J_t].request$
		\State $J \gets \texttt{consensus}(\mathbf{y}_t, \mathbf{z}_t, \mathbf{y}_k, \mathbf{z}_k)$
		\If{$Tasks[J] \in b_t$}
		\State $task \gets b_t.\texttt{pop()}$
		\EndIf
		\EndFor
		\EndWhile
		\State \Return $b_t$
		\EndFunction
	\end{algorithmic}
\end{algorithm}

\subsubsection{Algorithm 5}
Algorithm 5 illustrates the task unpacking and sorting for $agent_t$.
\begin{algorithm}
	\caption{task unpacking and sorting for $agent_t$}
	\textbf{Symbol}: \texttt{ungroup($b_t$, groupList)} \\
	\textbf{Input}: Bid-winning task queue $b_t$, group list \texttt{groupList} \\
	\textbf{Output}: ordered target location queue $p_i$
	\begin{algorithmic}[1]
		\Function{ungroup}{$b_t$, groupList}
		\State \texttt{bundle} $\gets$ \texttt{list()} 
		\For{each $id \in b_t$}
		\For{each task $e \in \texttt{groupList}[id]$}
		\State \texttt{bundle.add} $(task.start, task.end)$
		\EndFor
		\EndFor
		\State $p_i \gets \texttt{list()}$
		\While{\texttt{bundle} is not empty}
		\For{each $j \in \texttt{length(bundle)}$}
		\State $c_{ij} \gets s_{ij}(p_i)$
		\EndFor
		\State $J_i \gets \texttt{argmax}(c_{ij})$
		\State $n_i \gets \texttt{argmax}(s^2_{ij})$
		\State $p_i.\texttt{add}(\texttt{bundle}[J_i], n_i)$
		\EndWhile
		\State \Return $p_i$
		\EndFunction
	\end{algorithmic}
\end{algorithm}

\subsubsection{TABLE I}
TABLE I presents a complete consensus strategy provided by CBGA.
\begin{table}[h]
	\centering
	\caption{Action rules for task $j$ in Communication between agent$_k$ and agent$_t$.}
	\begin{tabular}{ccc}
		\hline
		\begin{tabular}[c]{@{}c@{}}agent$_k$\\ considers that\\ $z_{kj}$ is equal\end{tabular} &
		\begin{tabular}[c]{@{}c@{}}agent$_t$\\ considers that\\ $z_{kj}$ is equal\end{tabular} &
		\begin{tabular}[c]{@{}c@{}}Action of agent$_k$\end{tabular} \\
		\hline
		& $i$ & if $y_{kj} > y_{ij}$: \texttt{update} \\
		\cline{2-3}
		$k$& $k$ & \texttt{update} \\
		\cline{2-3}
		& $m \notin \{i,k\}$  & if $t_{km} > t_{im}$ or $y_{kj} > y_{ij}$: \texttt{update} \\
		\cline{2-3}
		& none& \texttt{update} \\
		\hline
		& $i$ & \texttt{leave} \\
		\cline{2-3}
		$i$ & $k$ & \texttt{reset} \\
		\cline{2-3}
		& $m \in \{i,k\}$ & if $t_{km} > t_{im}$: \texttt{reset} \\
		\cline{2-3}
		& none & \texttt{leave} \\
		\hline
		& $i$ & if $t_{km} > t_{im}$ and $y_{kj} > y_{ij}$: \texttt{update} \\
		\cline{2-3}
		& $k$ & if $t_{km} > t_{im}$: \texttt{update} else: \texttt{reset} \\
		\cline{2-3}
		$m \notin \{i,k\}$ & $m \in \{i,k\}$ & if $t_{km} > t_{im}$: \texttt{update} \\
		\cline{2-3}
		& $n \in \{i,k,m\}$ & \begin{tabular}[c]{@{}c@{}}if $t_{km} > t_{im}$ and $t_{kn} > t_{in}$: \texttt{update} \\
			if $t_{km} > t_{im}$ and $y_{kj} > y_{ij}$: \texttt{update} \\
			if $t_{kn} > t_{in}$ and $t_{im} > t_{km}$: \texttt{reset} \\ if $t_{km} > t_{im}$: \texttt{update} \end{tabular}\\
		\cline{2-3}
		& none & if $t_{km} > t_{im}$ :update \\
		\hline
		& $i$ & \texttt{leave} \\
		\cline{2-3}
		none& $k$ & \texttt{update}\\
		\cline{2-3}
		& $m \notin \{i,k\}$ & if $t_{km} > t_{im}$: \texttt{update} \\
		\cline{2-3}
		&none & \texttt{leave} \\
		\hline
	\end{tabular}\label{tab1}
\end{table}

\subsubsection{TABLE II}
The experimental design is summarized in TABLE II.
\begin{table}[htbp]
	\caption{Experimental Design Scheme}
	\label{table}
	\setlength{\tabcolsep}{3pt}
	\begin{tabular}{p{50pt}p{65pt}p{105pt}}
		\hline
		Experiment Number& 
		Experiment Objective& 
		Experiment Method \\
		\hline
		Experiment 1 &
		Test the performance of GCBHA in solving task allocation problems. &
		Compare the algorithm runtime and the total score of the solutions required for task allocation by GCBHA, CBGA, and nCAR under the conditions of varying numbers of tasks and agents, as well as whether the carrying capacity of the agents is considered. \\
		\hline    
		Experiment 2 &
		Evaluate the performance of GCBHA in resolving GMAPD issues, and verify the precision of the distance estimation method in warehouse scenarios. &
		Compare the performance of CENTRAL, CBGA, GCBHA, and TA-priority in solving GMAPD under different parameter conditions, and summarize the algorithm runtime, the length of the solution path, as well as the difference between the estimated path length and the actual path length. \\
		\hline
	\end{tabular}
	\label{tab2}
\end{table}

\subsubsection{TABLE III}
The detailed experimental parameter settings are shown in TABLE III.
\begin{table*}[htbp]
	\caption{Experimental Parameter Configuration}
	\centering
	\begin{tabular}{ccc}
		\hline
		
		Parameter Name	& Experiment 1	& Experiment 2 \\
		\hline
		algorithm	&GCBHA, CBGA, nCAR	& GCBHA, CBGA, CENTRAL, TA-priority \\
		$\lambda$	&0.1 &	0.1 \\
		type of map 	& Warehouse &	Warehouse \\
		size of map 	& "80×80"  &	"80×80" \\
		number of tasks and agents (ask, agent)	& (20,10), (50,10), (50,20), (100,20), (100,50), (200,50), (200,100)	&  (50,20), (100,20), (200,20), \\ (50,50), (100,50), (200,50) \\
		speed of small-scale agent &	1	&1 \\
		carrying capacity of small-scale agents	&100	& 100 \\
		speed of large-scale agent &	2	& 2 \\
		carrying capacity of large-scale agents &	200	& 200 \\
		percentage of large-scale agents &	10\%	& 10\% \\
		
		general task requirement &	[10,50] &	[10,50]\\
		large task requirement	& (200,300]	& (200,300] \\
		percentage of large-scale tasks	& 10\%	& 10\% \\
		group task requirement &50, 100&	50\\
		percentage of special tasks	& 10\%	&10\%\\
		task value and percentage of requirement	&10&	10\\
		number of experiments per parameter group	&10&	10\\
		\hline
	\end{tabular}
	\label{tab3}
\end{table*}

\end{document}